\documentclass[twocolumn,showpacs,amsmath,amssymb,prl,lengthcheck]{revtex4}
\usepackage{graphicx}
\usepackage{dcolumn}
\usepackage{bm}
\usepackage{psfig}

\newcommand{\be}{\begin{equation}}
\newcommand{\ee}{\end{equation}}
\newcommand{\ba}{\begin{array}}
\newcommand{\ea}{\end{array}}
\newcommand{\bqa}{\begin{eqnarray}}
\newcommand{\eqa}{\end{eqnarray}}

\begin{document}
\preprint{Draft-PRL}

\title{ \bf \boldmath Analyticity and the $N_c$ counting rule of $S$ matrix poles}

\author{ Z.~G.~Xiao and H.~Q.~Zheng
\vspace{0.2cm}\\
 Department of Physics, Peking University, Beijing
100871, P.~R. China}

\date{\today}

\begin{abstract}
{By studying $\pi\pi$ scattering amplitudes in the large $N_c$
limit, we clarify the $N_c$ dependence of the $S$ matrix pole
position. It is demonstrated that analyticity  and the $N_c$
counting rule exclude the existence of $S$ matrix poles with
${\cal M},\,\,\Gamma\sim O(1)$. Especially the properties of
$\sigma$ and $f_0(980)$ with respect to the $1/N_c$ expansion are
discussed. We point out that in general tetra-quark resonances do
not exist. }
\end{abstract}
\pacs{11.15.Pg, 11.55.Bq, 12.39.Fe, 14.40.Cs}
 \maketitle

Recently there are increased interests in investigating the nature
of the $f_0(600)$ or the $\sigma$ resonance, which is important
for a deeper  understanding  of spontaneous chiral symmetry
breaking of QCD. Also there are revived interests in recent
literature to search for exotic states, for example, the tetra
quark states appear in meson-meson scatterings. In this Letter we
devote to the study of these  problems, using techniques from the
$S$ matrix theory, low energy effective theory and large $N_c$
expansion.
 Considering the difficulty of the
problem, we will mainly confine ourselves in dealing with elastic
$\pi\pi$ scattering amplitudes.

We begin by noticing that,  for the partial wave elastic
scattering, the physical $S$ matrix can in general be factorized
as,~\cite{piK,zhou}
 \be\label{param}
S^{phy}=\prod_iS^{p,\,i}\cdot S^{cut}\ ,
 \ee
 where $S^{p,i}$ are
the simplest $S$ matrices characterizing  isolated singularities
of $S^{phy}$, that is, for virtual/bound states: $S(s)=(1\pm
i\rho(s)a )/(1\mp i\rho(s)a)$ and for a resonance located at $z_0$
($z_0^*$) on the second sheet: $S^R(s)=(M^2[z_0]-s+i\rho(s)s\,
G[z_0])/ (M^2[z_0]-s-i\rho(s)s\, G[z_0])$ where
 \bqa\label{z0depen}
 M^2[z_0]&=&{\rm
Re}[z_0] + {\rm Im}[z_0]\frac{{\rm
Im}[\sqrt{z_0(z_0-4m_\pi^2)}]}{{\rm
Re}[\sqrt{z_0(z_0-4m_\pi^2)}]}\ ,\nonumber\\
G[z_0]&=&\frac{{\rm Im}[z_0]}{{\rm Re}[\sqrt{z_0(z_0-4m_\pi^2)}]}\
. \eqa
 The functions $M^2[z_0]$
and $G[z_0]$ have the properties that if $\mathrm{ Im}[z_0]\neq 0$
then either $M^2[z_0]>4m_\pi^2$, $G[z_0]>0$,  or $M^2[z_0]<0$,
$G[z_0]<0$. These properties will be
 useful later. The pole
 mass and width are denoted as $z_0\equiv ({\cal
 M}+i\Gamma/2)^2$ in this Letter. For the reason which will become
 apparent later, parameters $M^2$ and $G$ (or $\mathrm{Re}[z_0]$, $\mathrm{Im}[z_0]$) more
 appropriately describe a resonance
 than $\cal M$ and $\Gamma$. For a narrow resonance
 located in the region detectable experimentally
we have approximately ${\cal M}=M$, $G=\Gamma/M$. It is also worth
noticing that the resonance and also the virtual state
contributions to the scattering length and phase shift are always
positive whereas the bound state pole contribution is always
negative.

The $S^{cut}$ contains only cuts which can be parameterized in the
following simple form,~\cite{zhou} \bqa\label{fS'}
 S^{cut}&=&e^{2i\rho f(s)}\ ,\\
 f(s)&=&\frac{s}{\pi}\int_{L}\frac{{\rm
 Im}_Lf(s')}{s'(s'-s)}ds'+\frac{s}{\pi}\int_{R}\frac{{\rm
 Im}_Rf(s')}{s'(s'-s)}ds'\nonumber\\
&\equiv&f_L(s)+f_R(s)\ . \label{fS} \eqa
 where $L=(-\infty,0]$ and $R$ denotes cuts at higher energies other than the
$2\pi$ elastic cut. It starts at 4$\pi$ threshold but to a good
approximation it starts at $4m_K^2$.
 The discontinuity $f$ obeys  the
following simple relation: \bqa\label{IMLR}
&&\mathrm{Im}_{L,R}f(s)=-{1\over
{2\rho(s)}}\log|S^{phy}(s)|\nonumber\\
&&=
-\frac{1}{4\rho}\log\left[1-4\rho\mathrm{Im}_{L,R}T+4\rho^2|T(s)|^2\right]\
.
\eqa%
Before proceeding it is important to notice that
Eqs.~(\ref{fS}) and (\ref{IMLR}) are obtained by assuming that the
partial wave amplitudes are analytic on the whole cut plane, which
can be derived, for example,  by assuming the validity of
Mandelstam representation for the full $T$ matrix
amplitude.
 Nevertheless
the validity of Mandelstam representation  goes beyond what is
rigorously established from field theory, and what is rigorously
established on analyticity of partial wave amplitudes is the
Lehmann--Martin domain of analyticity.~\cite{MC70} If
Eq.~(\ref{fS}) becomes invalid, it should be replaced by the
following expression:
\begin{eqnarray*}
f(s)&=&\frac{s}{2\pi
i}\int_{C}\frac{f_C(s')}{s'(s'-s)}ds'+\frac{s}{\pi}\int_{R}\frac{{\rm
 Im}_Rf(s')}{s'(s'-s)}ds'\ ,\\
 f_C&=&\frac{1}{2i\rho}\log\left(1+2i\rho T^{phy}\right)\ , \hspace{2.7cm}(4')
\end{eqnarray*}
where $C$ is the contour separating the Lehmann--Martin domain and
the region unknown. In the following we will  assume the validity
of Eq.~(\ref{fS}) but our major conclusions on the $N_c$
properties of resonance states depend very little on
Eq.~(\ref{fS}).

Starting from Eq.~(\ref{IMLR}), the first observation is, the
integrand of the left hand dispersion integral for function $f$ is
not allowed to make a chiral perturbation expansion, due to the
$1/\sqrt{s}$ singularity hidden in the relativistic kinematic
factor $\rho(s)$.  This phenomenon does not necessarily lead to
any profound impact on the validity of chiral expansions. This may
be best illustrated by the following example, the integral
 \bqa\label{example}
 && \frac{1}{2\pi}\int_0^\infty \frac{\log(1 +
\alpha^2/x)}{\sqrt{x}(1+x)}dx =\log(1+|\alpha|)\nonumber\\
&&= \alpha -\frac{\alpha^2}{2}+\frac{\alpha^3}{3}+\cdots\ ,
\,\,\,\,\,\, (\alpha>0)\eqa
 does not allow an expansion on the
integrand in powers of the coupling constant $\alpha$, but after
performing the integration it can still be expanded in powers of
$\alpha$ (though the expansion is not analytic at $\alpha=0$). A
related problem is the $N_c$ power counting of $f_L$. Since $T\sim
O(N_c^{-1})$ when making the large $N_c$ expansion one may naively
neglect the the term proportional to $|T|^2$ inside the logarithm
on the $r.h.s.$ of the second equality of Eq.~(\ref{IMLR}) since
it is $O(N_c^{-2})$, but Eq.~(\ref{example}) reveals that it will
come back and make a contribution at $O(N_c^{-1})$. On the other
side, $\chi$PT predicts $\mathrm{Im}_LT$ to be $O(N_c^{-2})$, but
at higher energies it is no longer true. We in fact have
$\mathrm{Im}_LT\sim O(N_c^{-1})$ from further left hand cuts
contributed by crossed channel resonance exchanges. To see this
recall that,~\cite{MMS}
 \bqa\label{FGPF}
&& \mathrm{Im}\,T_l^I(s)=\frac{[1+(-1)^{l+I}]}{s-4}\sum_{l'}\sum_{I'}(2l'+1)C_{II'}^{(st)}\nonumber\\
&&\times\int^{4-s}_4 dt
P_l(1+\frac{2t}{s-4})P_{l'}(1+\frac{2s}{t-4})\mathrm{Im}\,T^{I'}_{l'}(t)\
,\nonumber\\
 \eqa
which relates the nearby left hand cut to the  physical region
singularities. Using Eq.~(\ref{FGPF}) together with narrow
resonance approximation $\mathrm{Im}T(t)\sim\pi\sum\Gamma_i
M_i\delta(M_i^2-t)$ one finds indeed that the left cut is
$O(N_c^{-1})$.~\cite{MMS'}

 A natural way to avoid the problem of the expansion at $s=0$
  as mentioned above is to make instead a
threshold expansion  on $f$, $
 f=\sum_{n=0}^\infty f_n\frac{(s-4m_\pi^2)^n}{(m_\pi^{2})^n}\ .
$ The contributions from the left hand and right hand integral to
each coefficient will be denoted as $f_{Ln}$ and $f_{Rn}$,
respectively. In the large $N_c$ limit  $f_{L}$ can be written as:
 \be f_{L}(s) = \frac{s}{\pi}\int_{L}\frac{{\rm
 Im}_LT(s')}{s'(s'-s)}ds'- |T(0)|+ O(N_c^{-2})\ , \label{f0NcOrder}
\ee where the $O(N_c^{-2})$ part of  $\mathrm{Im}_LT$ and $T(0)$
are to be neglected. Here we will not attempt to calculate the
left cut explicitly. What is really important to us is that
$\mathrm{Im}_LT$ (and hence $f_{Ln}$) is $O(N_c^{-1})$, it
certainly can not be $O(1)$ since $T$ itself is $O(N_c^{-1})$.
Another important fact of Eq.~(\ref{f0NcOrder}) is that it makes
sense to make a low energy expansion to the integral, since the
integration starts effectively from $4m_\pi^2-M^2$ (where $M$ is
the mass of the lightest crossed channel resonance), rather than
from 0.

 In Eq.~(\ref{param}) only second sheet poles are explicitly parameterized and poles on
 other sheets are all hidden in $f_R$. For the latter,  it is
 naively expected to be
 also of $O(1/N_c^2)$.
  But since there are poles on sheets closely connected with the physical
  region
  approaching
  the upper half of the physical cut (for example in $\pi\pi$,
  $\bar KK$ couple channel system, narrow poles on the
  third sheet, but not on the fourth sheet).  In the large $N_c$ limit,
  the integration path will come over those poles and  pinched singularities will occur. The $N_c$
  order will also change in this case.
  When only these poles are considered, using parametrization \cite{collins}
  \bqa\label{inelasticT}T_{1n}=\frac1{\sqrt{\rho_1(s)\rho_n(s)}}
  \sum_{r}{ {\cal M}_r(\Gamma_{r1}\Gamma_{rn})^
  {\frac{1}{2}}\over {\cal M}_r^2-s-i{\cal M}_r \Gamma_r}+ C\, ,
  \eqa with $\Gamma_{rn}$ the partial width, $\Gamma_{r}$ the total width
  and $C$ the smooth background at most of $O(N_c^{-1})$,
  and taking, for example, $\Gamma_r\sim O(1/N_c)$, the $r.h.c.$ integral can be carried out:
  \bqa \label{intR}
  \frac{s}{\pi}\int_{R}\frac{{\rm Im}_Rf(s')}{s'(s'-s)}ds'
  =\sum_r\frac{G_rs}{{\cal M}_r^2-s} \, ,
  \eqa
  which holds in the large $N_c$ limit when $s<<M_r^2$ and
$G_r=\frac{\Gamma_r(1-\sqrt{1-\alpha_r})}
  {2\sqrt{({\cal M}_r^2-4m_\pi^2)}}\ $,
  where $\alpha_r={4\Gamma_{r1}(\Gamma_r-\Gamma_{r1})}/{\Gamma_r^2}$.
  $G_r$ is distinguished from $G$ of the second sheet resonances
by subscript $r$. Now $f_{R0}$, $f_{R1}$, $f_{R2}$ can be estimated,
  \bqa \label{fr}
  f_{R0}&=&\sum_r{4m_\pi^2 G_r\over {\cal
  M}_r^2-4m_\pi^2}\ ,\,\,\,  f_{R1}=\sum_r{{\cal M}_r^2m_\pi^2 G_r\over
  ({\cal M}_r^2-4m_\pi^2)^2}\, ,\nonumber \\
  f_{R2}&=&\sum_r{{\cal M}_r^2m_\pi^4 G_r\over ({\cal M}_r^2-4m_\pi^2)^3}\ ,
  \eqa
  where higher order terms of $1/N_c$ expansion are neglected.
  From Eq.~(\ref{intR}) we find that after integration the higher sheet poles contributions
  are really of $O(1/N_c)$. In the definition of $G_r$, since
  ${\cal M}_r>4m_K^2$ if neglecting the $4\pi$ cut, and by definition
  $\Gamma_r>\Gamma_{r1}$, $G_r$ is positive. From Eq.~(\ref{fr}) we can see that $f_{R0}$,
  $f_{R1}$, $f_{R2}$ are all positive (which
  can actually be directly obtained from positivity of Im$_RT$).
    It should be realized that if there exists a higher sheet pole
  with $\Gamma_r\sim O(1)$,
  $M^2_r\sim O(1)$, i.e., not approaching real axis, then
 such a pole does not enter into Eq.~(\ref{fr}). The effect of such a pole
can only be cancelled by a nearby zero (i.e., a pole on other
sheets). It  cannot be cancelled by a nearby pole on the same
sheet because the other
 pole has a negative norm (like the time-like component of the photon
 field) in order to make the cancellation take place and to  make the elastic $\pi\pi$ scattering  amplitude
  $O(1/N_c)$ (since they do not approach real axis, the net effect to Eq.~(\ref{fr}) after the cancellation
   is $1/N_c^2$ suppressed).
  However, a pole with negative norm causes
 the severe problem of negative probability and hence should not appear.
  Also it should be mentioned that here we can not exclude,
  by only looking at the $N_c$ order of $T_{\pi\pi\to\pi\pi}$
  amplitude,
  a higher sheet pole like,
$\Gamma\sim O(1)$, ${\cal M}\sim O(1)$ but $\Gamma_{r1}\sim
  O(1/N_c^2)$. The existence of  such  a resonance does not contradict the $N_c$ counting rule of
  $T_{\pi\pi\to \pi\pi}$. The problem may only be studied by analyzing the, for example,
   $T_{\bar KK\to \bar KK}$ amplitude.
Related discussions will be given later.

According to the conventional wisdom, the complete $S$ matrix
defined in Eq.~(\ref{param}) can be faithfully parameterized by
the low energy effective theory, i.e., $S^{phy}=S^{\chi PT}$, in a
limited low energy region on the complex $s$ plane. Implicitly the
above statement requires that there is a convergence radius for
the low energy theory which do not shrinks to zero for arbitrary
value of $N_c$. This is necessary because otherwise we can not
make any expansion. Also it requires that there is no bound state
pole for $\pi\pi$ scatterings in $N_c$ QCD, since inside the
validity domain, the physical spectrum as predicted by the low
energy theory should be respected.  The condition on the absence
of bound state pole is not absolutely necessary for our later
purpose, though it will simplify our discussion considerably .
What we will do in the following is to make a threshold expansion
to the $S$ matrices of resonance poles, $S^p\equiv\prod_i
S^{p,\,i}$, on the $r.h.s.$ of Eq.~(\ref{param}). For the present
purpose we recast Eq.~(\ref{param}) as,
 \be\label{match'}
  S^p(s)=S^{phy}(S^{cut})^{-1}=S^{phy}e^{-2i\rho\,f(s)}\ .
   \ee
   The matching between the $l.h.s.$ and the $r.h.s.$ of the above equation
can be performed at sufficiently small energies if we make
  a threshold as well as a chiral  expansion on $S^{phy}$, that is to
replace $S^{phy}$ on the $r.h.s.$ of the above
   equation by
$S^{\chi PT}$, for the latter we have the standard $N_c$ counting
rules~\cite{GL_NC}. It has been illustrated that one can make a
low energy expansion on $f(s)$ on the $r.h.s.$ of
Eq.~(\ref{match'}) as well, with each coefficients at most
$O(N_c^{-1})$. Therefore the matching between the $l.h.s.$ and the
$r.h.s.$ of Eq.~(\ref{match'}) can be done in the leading order of
$1/N_c$ expansion.
 $S^p(s)$
can be safely expanded at the $\pi\pi$ threshold, since for any
resonance pole not lying on the real axis there is always
$M^2[z_0]\neq 4m_\pi^2$. For simplicity of discussions we did not
include virtual poles, but no conclusion will be changed if they
are included.  It is straightforward to demonstrate the following
relation, \bqa\label{compare2} &&\frac{(S^p(s)-1)}{2i\rho(s)}=
\sum_i{4G_i m_\pi^2\over M_i^2
-4m_\pi^2}+\sum_i{G_iM_i^2\over(M_i^2-4m_\pi^2)^2}\nonumber\\&&\times
(s-4m_\pi^2) +\sum_i{G_iM_i^2\over(M_i^2-4m_\pi^2)^3}
(s-4m_\pi^2)^2+\cdots\nonumber\\&&+O(N_c^{-2})\ .
 \eqa
Every coefficient of the series on the $r.h.s.$ of the above
equation is $O(N_c^{-1})$. Since there are no bound state  poles
by assumption (virtual state poles are harmless), then the only
isolated singularities appeared here are the second sheet
resonances. Then according to the properties of $M^2_i$ and $G_i$,
every term in the first (and also the second, but not the third)
coefficient is positive. Therefore \textit{ the $N_c$ order of
each $G_i/(M_i^2-4m_\pi^2)$ can not be larger than -1}, since no
cancellation is possible due to the positivity of each term.
Particularly  we demonstrate here that there cannot be states
behaving like ${\cal M}\sim O(1)$, $\Gamma\sim O(1)$. Furthermore,
if $M_i^2$ is non-vanishing in the chiral limit, then we have
$G_i/M^2_i \sim O(1/N_c)$ or less.
 For these poles with $G_i/M^2_i \sim O(1/N_c)$,
from the $N_c$ dependence of the third term on the $r.h.s.$ of
Eq.~(\ref{compare2}), one concludes that there should be at least
one pole with $G\sim O(1/N_c)$, $M^2\sim O(1)$. Such a pole
corresponds to the normal resonance made of one quark and one
anti-quark when $M^2$ is positive, that is ${\cal M}\sim O(1)$,
$\Gamma\sim O(1/N_c)$. Such a result is not surprising at all
since it is the the standard $N_c$ counting rule for normal
mesons.~\cite{Witten} However, our derivation is  valuable since
the $S$ matrix pole's correspondence to the quark composites is
not totally clear. The virtual pole in the IJ=20 channel, located
at $s_0=m_\pi^6/(16\pi^2F_\pi^4)+O(m_\pi^8)$ on the $s$ plane,  is
a living counter example.~\cite{ang}  If we consider only ordinary
poles made of quarks and gluons, then ${\cal M}\sim O(1)$ as a
fact of wave function normalization and $\Gamma\sim O(1/N_c)$ or
less. The latter situation can not be excluded and may well happen
in nature. For example, a glueball's decay width to $\pi\pi$ is
$O(1/N^2_c)$.

The matching up to and including $(s-4m^2_\pi)^2$ terms leads to
the following three equations in the leading order of $O(1/N_c)$
expansion:
 \bqa
 \sum_{n=i,r}{4G_{n} m_\pi^2\over M_{n}^2
-4m_\pi^2}&=&T_0^{\chi PT}-f_{L0}^{}\ ,\label{matching}\\
\sum_{n=i,r}{G_{n}M_{n}^2m_\pi^2\over(M_{n}^2-4m_\pi^2)^2}&=&T_1^{\chi
PT}
-f_{L1}^{}\ ,\label{matching2}\\
\sum_{n=i,r}{G_{n}M_{n}^2m_\pi^4\over(M_{n}-4m_\pi^2)^3}&=&T_2^{\chi
PT}-f_{L2}^{}\ ,\label{matching3}
 \eqa
where we have already replaced the partial wave $T$ matrix on the
$r.h.s$ of above equations by 1--loop chiral perturbation
amplitudes: $T(s)=T^{\chi PT}(s)+O(p^6).$ Notice that in above
equations the subscripts in $T^{\chi PT}$ and $f_{L}$ imply the
order of threshold expansion. It is not difficult to check using
the results of chiral amplitudes that
  Eq.~(\ref{matching}) and Eq.~(\ref{matching2}) are degenerate
in the chiral limit. However, except for being helpful in
determining the $N_c$ counting of resonances, the equations
(\ref{matching}) -- (\ref{matching3}) are of little use if one
does not know how to calculate  those $f_{Ln}$ coefficients. Here
we only point out that it is a good speculation to neglect
numerically those crossed channel effects in the IJ=11 channel,
since the left cut contribution is tiny in this
channel~\cite{zhou}. Then Eqs.~(\ref{matching}) --
(\ref{matching3}) read,
 \bqa\label{rhosr}
&&\sum_i\frac{G_{v,i}}{M_{v,i}^2}+\sum_r\frac{G_{v,r}}{M_{v,r}^2}=\frac{1}{96\pi F_\pi^2}\ ,\\
&&\sum_i\frac{G_{v,i}}{M_{v,i}^4}+\sum_r\frac{G_{v,r}}{M_{v,r}^4}=-\frac{L_3}{24\pi F_\pi^4}\
.\label{rhosr2}
 \eqa
Every parameter in above equations is understood as the
corresponding value in the large $N_c$ and chiral limit. Good
agreement are found between two sides of the above two equations.
Actually when neglecting the higher resonances (which are very
small numerically) the Eq.~(\ref{rhosr}) reproduces the well known
KSFR relation. Left cut contributions in the IJ=20 and 00 channels
are large, therefore it is not correct to neglect them at all.
What we would like to emphasize here is that the $\sigma$ pole
must behave as $G/M^2\sim O(N_c^{-1})$ if it contributes to
Eq.~(\ref{matching}) and/or (\ref{matching2}) (or equivalently
speaking, it contributes to $F_\pi$).  It should further behave as
$G\sim O(N_c^{-1})$, $M^2\sim O(1)$, if it also contributes to
Eq.~(\ref{matching3}) (i.e., it contributes to the LECs, the $L_i$
parameters~\cite{GL_NC}). Related model dependent discussions  on
$\sigma$ trajectory may  be found in Ref.~\cite{sunzx}.

When obtaining the $N_c$ counting rule for $S$ matrix poles, we
rely on the  analyticity property of the partial wave $S$ matrix.
One of the most important result is that resonances with $M^2\sim
O(1)$ and $G\sim O(1)$ do {\it not} exist. The results do not
depend much on whether we have analyticity on the whole cut plane.
It is obtained based on two conditions: 1) all the $S$ matrix
poles appeared on the $l.h.s$ of Eq.~(\ref{matching}) have the
same sign, because  they  are all located in the Lehmann--Martin
domain of analyticity; 2) the fact that the $r.h.s$ of
Eq.~(\ref{matching}) is $O(N_c^{-1})$. The latter condition
follows from the fact that the $T$ matrix itself is $O(N_c^{-1})$,
and it remains to be true even if using Eq.~(4$'$) instead of
Eq.~(4). It is of little interests to discuss the  $N_c$
dependence of poles located outside the Lehmann--Martin domain, if
there are any.

The above discussions are only limited to $\pi\pi$ scatterings.
However we believe the picture should also hold for any two
pseudo-Goldstone boson scatterings, since the only difference
comes from kinematics which should not waver the $N_c$ counting
rule. Taking $\pi\pi$, $\bar KK$ couple channel system for
example,
 the analytic continuation of partial wave $S$ matrices
 on different sheets are
 \bqa \label{Ss}&&S^{II}=\left(
\begin{array}{cc} {1\over S_{11}} & {i S_{12}\over S_{11}}\\
{i S_{12}\over S_{11}} &{{\rm det} S \over
S_{11}}\end{array}\right)\ ,\,\,\,S^{III}=\left(
\begin{array}{cc} {S_{22}\over {\rm det}S} & {-S_{12}\over {\rm det} S} \\
{-S_{12}\over {\rm det}S} &{S_{11} \over {\rm
det}S}\end{array}\right)\ ,\nonumber\\
&&S^{IV}=\left(\begin{array}{cc}{{\rm det}S\over S_{22}}  & -{i
S_{12}\over S_{22}} \\ -{i S_{12}\over S_{22}} & {1 \over S_{22}}
\end{array}\right)\ .
  \eqa
 From these expressions we realize
 that
  a third sheet pole in $S_{\pi\pi\to\pi\pi}$ which resides on
the complex $s$ plane when $N_c$ large may only be cancelled by a
nearby zero (a fourth sheet pole), i.e., they annihilate at
$N_c=\infty$. But a third sheet pole in $T_{\pi\pi\to \pi\pi}$ is
also a third sheet pole in $T_{\bar KK\to \bar KK}$. The wrongful
$N_c$ order of such a pole in the $T_{\bar KK\to \bar KK}$
amplitude may only be cancelled by a second sheet pole. However,
the cancellation is impossible to occur, since we have
demonstrated that no second sheet pole can reside on the complex s
plane except real axis in the large $N_c$ limit. Therefore we have
just argued that no partial width of any resonance pole can be
$O(1)$. The use of the analyticity condition is crucial in
obtaining the $N_c$ counting rule of meson resonances. In fact one
can construct a unitary amplitude with correct $N_c$ counting rule
but violating the analyticity condition. In such an example, the
second sheet pole resides on the complex $s$ plane when $N_c$ is
large, but it annihilates with a first sheet pole when
$N_c=\infty$.

However, in the $\pi\pi$, $\bar KK$ couple channel system, there
exists the $f_0(980)$ state which may be interpreted as a $\bar
KK$ molecular bound state. 
If it is indeed the case, then our previous discussions have to be
reexamined since some of our results obtained depend on the
hypothesis of the non-existence of bound state. If $f_0(980)$ is a
$\bar KK$ bound state, it has $O(1)$ coupling to $\bar KK$ and
$O(1/N_c)$ coupling to $\pi\pi$. Therefore it is a good
approximation to neglect  the $\pi\pi$ channel first.  In such a
case when there exists a bound state  it can still be demonstrated
that the cancellation between a bound state and resonance is
impossible at the level of $O(1)$. The only possibility to restore
the correct $N_c$ order of the scattering amplitude is to allow an
accompanying virtual state in association with the bound state. If
both $a_b$ and $a_v$ $\sim O(1)$, but $a_b-a_v\sim O(1/N_c)$, then
a correct $N_c$ counting for the $T$ matrix can be made. The
cancellation works because, \bqa S={(1-i \rho a_b)(1+i\rho
a_v)\over(1-i \rho a_b)(1+i\rho a_v) } ={ \frac{ 4 a_b a_v}{1+a_b
a_v}-s +i \rho s\frac{a_b-a_v}{1+a_ba_v}\over \frac{ 4 a_b
a_v}{1+a_b a_v} -s -i \rho
 s\frac{a_b-a_v}{1+a_ba_v}}\ ,\nonumber\\
\eqa i.e., the net effect after the cancellation is very much like
a normal resonance, but with a mass below the $\bar KK$ threshold,
$M^2=4 a_b a_v/(1+a_b a_v)<4m_K^2$.
 When coupling to $\pi\pi$ channel is
opening, the bound state becomes a narrow second sheet pole (and
hence described as the $f_0(980)$), and the virtual state becomes
a 3rd sheet  pole. Such a scenario is allowed within the present
scheme, and there exists evidence that the data are better
described by two poles near the $\bar KK$
threshold~\cite{MP93markushin97}. The discussions made above may
be used to argue the non-existence of tetra-quark
 states,~\cite{talkmenu04} except the  bound/virtual state scenario just mentioned.

To conclude, with the aid of analyticity condition we observe
 that any $S$ matrix pole trajectory on the $s$ plane obtained by increasing
 $N_c$ can only behave in one of
the following three ways: 1) always remains on the real axis; 2)
approaching real axis when $N_c\to \infty$; 3) moving to $\infty$
when $N_c\to\infty$. It is not totally clear whether the $\sigma$
follows the second or the third trajectory though the second one
is prefered. From the above observation we also conclude that in
general tetra quark states  do not exist.

 {\it Acknowledgement} We  would like to thank
F.~Kleefeld, C.~Liu, G. Rupp and S.~L.~Zhu for helpful
discussions. We also thank Z.~X.~Sun and L.~Y.~Xiao for
assistance.
 This work is support in part by China National
 Nature Science Foundations under contract number 10491306.


\begin{thebibliography}{99}
\bibitem{MMS'}The way we make the observation may not be   totally
unambiguous. Because rigorously speaking the Eq.~(\ref{FGPF}) only
works in a limited range of $s$.~\cite{MMS}
\end{thebibliography}
\end{document}